# Cooperative Behavior Cascades in Human Social Networks

James H. Fowler[1][*], Nicholas A. Christakis[2]

[1]Political Science Department, University of California, San Diego, La Jolla, CA  92103, USA

[2]Harvard Faculty of Arts and Sciences and Harvard Medical School, Boston, MA 02115, USA

∗ To whom correspondence should be addressed, email: jhfowler@ucsd.edu, phone: 858-534-6807.

Theoretical models suggest that social networks influence the evolution of cooperation, but to date there have been few experimental studies. Observational data suggest that a wide variety of behaviors may spread in human social networks, but subjects in such studies can choose to befriend people with similar behaviors, posing difficulty for causal inference. Here, we exploit a seminal set of laboratory experiments that originally showed that voluntary costly punishment can help sustain cooperation. In these experiments, subjects were randomly assigned to a sequence of different groups in order to play a series of single-shot public goods games with strangers; this feature allowed us to draw networks of interactions to explore how cooperative and uncooperative behavior spreads from person to person to person. We show that, in both an ordinary public goods game and in a public goods game with punishment, focal individuals are influenced by fellow group members' contribution behavior in future interactions with other individuals who were not a party to the initial interaction. Furthermore, this influence persists for multiple periods and spreads up to three degrees of separation (from person to person to person to person). The results suggest that each additional contribution a subject makes to the public good in the first period is tripled over the course of the experiment by other subjects who are directly or indirectly influenced to contribute more as a consequence. These are the first results to show experimentally that cooperative behavior cascades in human social networks.

\body

**Introduction**

Scholars studying the evolution of cooperation in humans have recently turned their attention to the role of social networks in structuring human interactions (*1–10*). Interacting with others in large populations without structure greatly reduces the likelihood of cooperation (*11*), but in a fixed social network, cooperation can evolve as a consequence of repeated interactions because of "social viscosity," even in the absence of reputation effects or strategic complexity (*1–2*). Different network structures can speed or slow selection, and, in some cases, they can completely determine the outcome of a frequency-dependent selection process (*3*). Heterogeneity in the interaction topology can improve prospects for cooperation (*4*), and adaptive selection of network ties by individuals on evolving graphs can also influence the evolution of behavioral types (*5–7*).

However, this theoretical literature has not explored whether cooperative behavior actually spreads across ties in human social networks, and recent experimental work has tended to focus on coordination rather than cooperation (*12–13*). A growing number of observational studies suggest that diverse phenomena can spread from person to person to person, including obesity (*14*), happiness (*15*), ideas (*16*) and many other behaviors and affective states (*17–20*). But causal effects are difficult to extract from observational network studies because similarity in observed attributes among connected individuals may result from homophily, the tendency to connect to others who exhibit similar traits or behaviors (*21*). For example, past work has shown that people who engage in acts of altruism tend to befriend others who do the same (*22*). Causal effects are also difficult to extract from observational studies because associations between connected individuals might be due to shared exposure to contextual factors that simultaneously engender various behaviors (including cooperation) in both parties.

Experimental studies can overcome these problems via random assignment of interactions in a controlled fashion. For example, a recent field experiment (*23*) showed that a door-to-door canvasser who encourages a resident to vote influences not only the person who answers the door but a second person in the household as well, even though there was no direct contact between the second person and the canvasser. But such studies experimentally showing person to person-to-person effects are rare. Prior experimental work on spreading processes in networks has focused primarily on direct person-to-person effects – for example with respect to the dyadic spread of studiousness (*24*), positive moods (*25–26*), and weight loss (*27*).

Not everything spreads between connected individuals, of course, and not everything that spreads does so by the same mechanism. For example, while the spread of emotional states, smiling, or yawning may be rooted in fundamental neurophysiological processes (*28*), the spread of behaviors may arise from the spread of social norms or from other psychosocial processes, such as various types of innate mimicry (*29*). In the particular case of cooperation or altruistic behavior, it is well known that one person's altruism towards another can elicit reciprocal altruism in repeated paired interactions (direct reciprocity) (*30*) and also in groups (*31–32*). Indeed, many individuals are "conditional cooperators" who give more if others give more and who are influenced in their interactions with others by what the others are doing during the interaction (*33–34*). And it is also well known that reputation mechanisms which provide information about a person's past behavior can help to sustain cooperation (indirect reciprocity) (*35*).

However, no work so far has considered the possibility that witnessing cooperative or uncooperative behavior might, by various mechanisms, including innate mimicry, promote changes in cooperative behavior that might then be transmitted across social network ties to others who were *not* part of the original interaction. That is, quite distinct from prior work, we are concerned with whether such behavior can create cascades of similar cooperative or uncooperative behavior in others, spreading from person to person to person, even when reputations are unknown and reciprocity is not possible. If so, it suggests that social contagion may also play an important role in the evolution of cooperation.

To study the spread of cooperative and uncooperative behavior in human social networks, we analyzed a set of previously published public goods game experiments (*36*). In these experiments, subjects were placed into groups of four, and each subject was given 20 money units (MUs). They then had to decide how many MUs (between 0 and 20) to keep or contribute to a group project. Each MU contributed to the group project would yield 0.4 MUs for each of the four group members. This set-up allows us to study cooperative behavior because each MU contributed is costly to the individual (0.4MU – 1MU = –0.6MU) but beneficial to the group (4 x 0.4MU –1MU = +0.6MU). If each group member keeps all MUs privately, they will each earn 20MUs, and if each group member makes the maximum contribution of 20MUs to the group project, they will each earn 32MUs. But in spite of this opportunity to improve group outcomes, each individual can always earn more by contributing less.

In the basic public goods game analyzed here, subjects made contribution decisions without initially seeing their fellow group members' decisions, but all contributions were revealed to each group member at the conclusion of the game, along with payoff outcomes. In an alternate version, subjects played an identical public goods game, but, after viewing their fellow group members' contributions, they were allowed to spend up to 10MUs to "punish" each of the other group members. Each 1MU spent reduced the target's income by 3MUs. In both versions of the experiment, subjects participated in a total of six repetitions ("periods") of the public goods game. This allowed the researchers to show that contributions tend to decline in the basic public goods game, and increase in the public goods game with punishment (*36*).

Importantly, to distinguish the effect of punishment on cooperation from the effect of direct reciprocity (*30*), indirect reciprocity (*35*), reputation (*37*), and costly signaling (*38*), the research design enforced strict anonymity between subjects, and no subject was ever paired with any other subject more than once. As shown in Fig.1, this feature of the experimental design allows us to construct networks of interactions where each connection is defined by the ability of one subject to observe the contribution behavior of another subject in the preceding period (because they were in the same group). Since random assignment rules out processes like homophily and contextual effects, a significant association in the public goods contributions of directly connected individuals suggests that one subject's cooperative or uncooperative behavior causally influences another person's behavior during his interactions with other, new subjects in the following period. Moreover, we can analyze associations between *indirectly* connected individuals to see whether such effects spread from person to person to person. For example, subject F may influence subject E who in turn influences subject A (see Fig.1), even though A did not interact with F or observe F's behavior. The mechanism of the effect of F upon A, of course, occurs *via* the effect of F on E and the subsequent effect of E on A.

**Results**

A summary of the raw data (Fig. 2) shows that, indeed, future contributions by focal individuals ("egos") are significantly related to the amount contributed by each group member with whom the ego interacted in the previous period ("alters") in both the basic public goods game and the public goods game with punishment. However, the raw relationship does not take into account constraints on the amount subjects can give (specifically, many subjects choose to contribute the minimum or maximum possible amounts). Nor does it take into account the fact that there are multiple observations for each ego and each alter within and between periods. We therefore use an interval regression technique with clustered standard errors (see Methods) to estimate the size of the causal effect of one subject's contribution behavior on another. We focus on the effect of each alter independently rather than each group of alters because it allows us to take into account ceiling and floor effects that occur at the individual level (see Methods). It also helps us conceptually to identify the spillover effect of a single individual on the people he or she is connected to, as well as the way this effect can subsequently spread to others in the interaction network.

The results show that alters significantly influence egos' behavior, both directly and indirectly (Fig. 3). For each 1MU contributed by an alter, ego contributes an additional 0.19 MUs (95% C.I. 0.14 to 0.24, $p$<0.0001) in the next period in the basic public goods game and an additional 0.18MUs (0.14 to 0.21, $p$<0.0001) in the public goods game with punishment. Note that these results summarize the spread of both cooperative and uncooperative behavior: alters who give less influence egos to give less and alters who give more influence egos to give more.

Remarkably, even though egos do not observe the contributions of their alters' alters (two degrees of separation), they too significantly affect ego's contribution decisions. Each 1MU contributed by alter's alter increases ego contributions by an additional 0.07 MUs (0.03 to 0.10, $p$=0.0005) two periods later in the public goods game, and by an additional 0.05 MUs (0.03 to 0.08, $p$=0.0001) in the public goods game with punishment. These effects reflect the *indirect* traces of an individual's actions on others via a chain of *direct* pairwise effects. Indeed, as expected given the experimental set-up, a Sobel test (*39*) shows that the effect of alter's alter on ego is *mediated* by the indirect effect that spreads from alter's alter to alter to ego (see SI). Furthermore, in the public goods game with punishment, we find evidence that cooperative behavior spreads one degree farther, up to three degrees of separation. Each 1MU contributed by an alter's alter's alter increases ego's contribution by 0.06 MUs (0.02 to 0.09, $p$=0.001).

As a separate matter, many of these direct and indirect effects endure, influencing ego's behavior long after the initial period of influence (Fig. 4). For example, F may influence E to give more not only in the following period when E interacts with A, I, and M, but also in the period after that when E interacts with K, P, and T (see Fig. 1), and in other future periods as well. In the basic public goods game, each 1MU contributed by the alter causes ego to contribute an additional 0.15MUs (95% C.I. 0.09 to 0.21, $p<0.0001$) two periods later, 0.08MUs (0.00 to 0.16, $p=0.04$) three periods later, 0.17MUs (0.07 to 0.27, $p=0.001$) four periods later, and 0.17MUs (0.00 to 0.33, $p=0.05$) five periods later. In the public goods game with punishment, each MU contributed by the alter causes ego to contribute an additional 0.15MUs (0.10 to 0.19, $p<0.0001$) two periods later, 0.11MUs (0.05 to 0.17, $p=0.0001$) three periods later, 0.14MUs (0.07 to 0.21, $p=0.0001$) four periods later, and 0.14MUs (0.03 to 0.25, $p=0.02$) five periods later. The effects of alters' alters decision (and not just the effect of alters' decision) also persist in the basic public goods game after four periods (0.07MUs; 0.01 to 0.13, $p=0.02$) and the public goods game with punishment after three periods (0.14MUs; 0.03 to 0.25, $p=0.007$).

To measure the overall size of these cooperative behavior cascades (see Fig. 5 for a hypothetical example), we focus only on those effects with $p$ values less than 0.05 in Figs. 3-4. In the basic public goods game, if a subject increased her contribution by an additional 1MU in period 1, it would *directly* cause the 3 other subjects in her group to increase their total contributions by 1.8MUs (95% C.I. 1.3 to 2.3) over the next 4 periods. It would also *indirectly* cause 9 other subjects to increase their total contributions by 1.2MUs (0.9 to 1.5) in periods 3 and 5, for an overall increase of 3MUs (2.4 to 3.6). In the public goods game with punishment, the direct increase in contributions would be 2.1MUs (1.6 to 2.7) over the next 5 periods and the indirect increase would be 0.9MUs (0.7 to 1.0) in periods 3 and 4, also totaling 3MUs (2.4 to 3.6). The reverse is also true, and each reduction in one's contribution in the first period can generate a cascade of *uncooperative* behavior through parts of the network subsequently. Yet, this exercise suggests that, overall, over the course of each version of the public goods game, the spread of cooperative behavior through the network approximately triples each additional contribution made in the first period, at least in a network of this particular, experimentally controlled structure.

Finally, we note that in the public goods game with punishment, alters' alters may also influence ego via their punishment behavior. F might punish E, causing E to contribute more after one period, which causes A to contribute more after two periods. To test this hypothesis, we regressed ego's contribution on the punishment alters received from others two periods ago (see SI). And, indeed, we find that punishment can spur cooperative cascades as well. Each punishment point alter's alter gives to alter increases *ego's* contribution two rounds later by 0.13MUs (95% C.I. 0.02 to 0.23, $p$=0.02). However, the effect does not appear to spread any further: the relationship between alter's alter's alter's punishment behavior and ego's contribution three periods later is insignificant ($p$=0.25). We also failed to find any evidence of spreading punishment behavior *per se*; the association between punishment received in the previous round and punishment given in the current round was not significant ($p$=0.83, see SI).

**Discussion**

It is often supposed that individuals in experiments like the one described here would seek to selfishly maximize their own payoffs. However, it is worth reiterating that most of our subjects violated this supposition: since all interactions are single-shot, the equilibrium prediction is to contribute nothing and to pay nothing to punish non-contributors, yet this was not what the subjects did (*36*). One mechanism that may underlie such deviations from "rational" action appears to be mimicry: when subjects copy the cooperative behavior of others with whom they interact, it causes them to deviate even more from rational self-interest and may help to reinforce this behavior.

Observational studies suggest that behaviors, knowledge, and emotions spread between people with personal social ties (*14–20*). Of course, people can be influenced by strangers too – for example, in catching a germ, following a worn path on a field, imitating a smile, adopting a fashion, or responding to road rage. In this experiment, relationships were anonymous and contact was not sustained. Nevertheless, there was real interaction, and people observed each others' behavior in the setting of a game that they cared about. A consistent explanation for both the experimental investigations and the observational studies is that people mimic the behavior they observe, and that this can cause behaviors to spread from person to person to person. If anything, it seems likely that people who are willing to copy strangers' behavior may be even more likely to copy similar behavior in friends in real world settings.

Although an act of punishment, like a contribution, can initiate a cooperative cascade, we found minimal differences in the spread of influence between the basic public goods game and the public goods game with punishment. This suggests that the existence of punishment does not fundamentally alter network dynamics: punishment may not enhance or facilitate the *spread* of cooperation per se. The reason for this may be that ego only observes whether alter is cooperating, not the motivations that alter has for behaving in a particular way, nor the prior history of interactions that alter himself had that may be prompting a particular behavior. However, punishment clearly has a direct effect on contributions, and the network process we describe may help to magnify the *indirect* effect of punishment. Thus, behavioral cascades may be a crucial part of an explanation for how small changes in human institutions (such as informal norms or formal rules about punishment) can yield large changes in a group's behavior. This is all the more remarkable since we found no evidence that punishment itself spreads.

This multiplier effect also suggests that behavioral imitation and inter-personal spread may be an important factor in the evolution of cooperation in humans. For example, cascades of cooperative (or non-cooperative) behavior can promote coordination on a particular strategy, which may decrease within-group variance. At the same time, the path-dependent nature of this process within each group may tend to increase between-group variance. In a population with structured groups, both of these effects work in favor of the emergence of altruism (*40*). Cascades may also help mitigate the negative effect of group size on cooperation (*11,41*) because they reduce the number of independent entities in a population, effectively increasing the size of groups in which public goods can be maintained via self-interest. Evolutionary game theorists should therefore consider the possibility that behavioral imitation itself may have coevolved with both cooperation and the emergence of social networks.

Such models might perhaps also help to explain the influence of genetic variation on social network structure (*42*). Egocentric network characteristics like network centrality can make some individuals more susceptible than others to contagions with negative outcomes for fitness (like germs, misinformation, and violence). But the results here suggest that one fundamental justification for the existence of elaborate social ties in the form of social networks may be that they may allow humans to benefit from the actions of widely distributed others, and may also allow them to spread beneficial strategies widely enough to benefit others on whom they depend. Genetic variation in social network position suggests that networks may influence reproduction or survival via a frequency-dependent selection process or rapid environmental variation (relative to evolution); in either case, given that cooperation itself also appears to have a genetic basis (*43*) it makes sense to think about how cooperation may play a role.

Finally, these results provide experimental support for a conjecture about human social networks. To explain a variety of observational data that show behavior in social networks is correlated up to three degrees of separation, but rarely farther, a "three degrees of influence rule" (*44*) has been described which suggests that 1) behavior can spread from person to person to person to person via a diverse set of mechanisms, subject to certain constraints, and 2) as a result, each person in a network can influence dozens or even hundreds of people, some of whom he or she does not know and has not met. The present results show experimentally that such cascades can occur in a controlled environment where people are making decisions about giving to others. Other researchers have shown that giving behavior can spread from person to person in natural settings, whether it is workplace donations to charity (*45*) or the decision to donate organs (*46*). However, it is an open question whether such "pay it forward" behavior spreads more widely from person to person to person in natural human networks.

## Materials and Methods

The procedures for implementing the public goods game experiments for the 240 subjects analyzed in this report have been described elsewhere (*36*). Fig. 1 in the main text illustrates that the requirement in these experiments that no two subjects meet each other twice ensures that any ego who is *directly* connected to an alter (one degree of separation) cannot also be connected *indirectly* to the same subject by two degrees of separation (an alter's alter). It also ensures there are no redundant paths at one and two degrees of separation and no subject can be connected to herself by two degrees. However, at three and four degrees of separation, such combinations are possible, so we remove from the analysis all self-connections and all redundant paths, and we keep just one observation from among those with the shortest path length (smallest degree of separation). For example, if at period *t* subject B is subject A's alter's alter's alter (3 degrees) via two paths and also his alter's alter's alter's alter (4 degrees) via five paths, for the purpose of analysis, we assign a single, randomly chosen observation for this pair to the data in which subject B's contribution behavior depends on subject A's behavior at $t$-3.

To analyze ego contribution behavior, we use interval regression (also known as "Tobit" regression), which is typical in the literature on public goods games (*31*). This type of regression model treats responses at the minimum (0MUs) and maximum (20MUs) as censored. Past work has shown that applying ordinary least squares regression to data like this yields inconsistent results (slope coefficients are biased towards zero and intercepts are biased away from zero), while interval regression yields consistent results (*47*). However, the coefficients in interval regression apply to the latent outcome variable (what subjects would do if they were not constrained) rather than the observed outcome variable (what subjects actually do).

To estimate the influence of one subject's contribution on another's contribution, we include in these regression models the alter's contribution in the period $t-s$, where $s$ is the degree of separation (alter: $s = 1$, alter's alter: $s = 2$, and so on). To control for serial correlation, we also include ego's contribution in the period $t-s$; alternative specifications that add additional lags (see SI) generate identical results. To control for period effects, we include an indicator variable for all but one of the periods in which ego contributions were observed. To control for multiple observations of the ego and the alter, we use Huber-White sandwich errors that account for errors clustered on each ego and each alter. As a robustness check, we examined whether the effect of alter on ego varies depending on whether alter's contribution is high or low (it does not). We also included the other two group members' contributions as a control variable, and this did not change the results either.

We further replicated all results treating the group contribution as the unit of analysis rather than the alter's contribution. And when we analyzed the effect of others' contributions on alter's influence over ego, we found that alter's influence remained significant under all conditions, which suggests that analysis at the individual level rather than the group level is appropriate (see SI).

We emphasize that all activity in the experiments was completely anonymous. Group composition changed randomly every period so that no one played with the same person more than once. The subjects were ignorant of other players' experimental history; neither past payoffs nor past decisions were known. Different group composition each period and the absence of any history of play ensured that subjects could neither develop reputations nor target other subjects for revenge.

**Acknowledgments**

Supported by grants from the National Institute on Aging (NIH P-01 AG031093 and P30 AG034420) and by a Pioneer Grant from the Robert Wood Johnson Foundation.

## References

1. Nowak MA (2006) Five rules for the evolution of cooperation. *Science* 314: 1560–1563

2. Ohtsuki H, Hauert C, Lieberman E, Nowak MA (2006) A simple rule for the evolution of cooperation on graphs and social networks. *Nature* 441: 502–505.

3. Lieberman E, Hauert C, Nowak MA (2005) Evolutionary dynamics on graphs. *Nature* 433: 312–316.

4. Santos FC, Santos MD, Pacheco JM (2008) Social diversity promotes the emergence of cooperation in public goods games. *Nature* 454: 213–216.

5. Skyrms B, Pemantle R (2000) A dynamic model of social network formation. *PNAS* 97: 9340–9346.

6. Pacheco JM, Traulsen A, Nowak MA (2006) Coevolution of strategy and structure in complex networks with dynamical linking. *Physical Review Letters* 97: 258103

7. Santos FC, Pacheco JM, Lenaerts T (2006) Cooperation prevails when individuals adjust their social ties. *PLOS Computational Biology* 2: 1284–1291.

8. Saavedra S, Reed-Tsochas F, Uzzi B (2009) A simple model of bipartite cooperation for ecological and organizational networks. *Nature* 457: 463–466.

9. Taylor PD, Day T, Wild G (2007) Evolution of cooperation in a finite homogeneous graph. *Nature* 447: 469–472.

10. Szabo G, Fath G (2007) Evolutionary games on graphs. *Phys Reports* 446: 97–216.

11. Enquist M, Leimar O (1993) The evolution of cooperation in mobile organisms. *Animal Behaviour* 45: 747–757.

12. Kearns M, Suri S, Montfort N (2006) An experimental study of the coloring problem on human subject networks. *Science* 313: 824–827.

13. Kearns M, Judd S, Tan J, Wortman J (2009) Behavioral experiments on biased voting in networks. *PNAS* 106: 1347–1352.

14. Christakis NA, Fowler JH (2007) The spread of obesity in a large social network over 32 years. *New England Journal of Medicine* 357: 370–379.

15. Fowler JH, Christakis NA (2008) The dynamic spread of happiness in a large social network. *Bristish Medical Journal* 337: a2338.

16. Singh J. (2005) Collaborative networks as determinants of knowledge diffusion patterns. *Management Science* 51: 756–770.

17. Christakis NA, Fowler JH (2008) The collective dynamics of smoking in a large social network. *New England Journal of Medicine* 358: 2249–58.

18. Cacioppo JT, Fowler JH, Christakis NA (2009) Alone in the crowd: The structure and spread of loneliness in a large social network. *Journal of Personality and Social Psychology* 97: 977–991.

19. Rosenquist JN, Murabito J, Fowler, JH, Christakis NA (in press) The spread of alcohol consumption behavior in a large social network. *Annals of Internal Medicine*.

20. Rosenquist JN, Fowler, JH, Christakis NA (in press) Social network determinants of depression. *Molecular Psychiatry*.

21. McPherson M, Smith-Lovin L, Cook JM (2001) Birds of a feather: Homophily in social networks. *Annual Review of Sociology* 27: 415–444.

22. Leider S, Mobius MM, Rosenblat T, Do QA (2009) Directed altruism and enforced reciprocity in social networks. *Quarterly Journal of Economics* 124: 1815–1851.

23. Nickerson DW (2008) Is Voting contagious? Evidence from two field experiments. *American Political Science Review* 102: 49–57.

24. Sacerdote B (2001) Peer effects with random assignment: Results for Dartmouth roommates. *Quarterly Journal of Economics* 116: 681–704.

25. Pugh SD (2001) Service with a smile: emotional contagion in the service encounter. *Acad Manag J* 44: 1018–27.

26. Howes MJ, Hokanson JE, Lowenstein DA (1985) Induction of depressive affect after prolonged exposure to a mildly depressed individual. *J Pers Soc Psychol* 49: 1110–13.

27. Gorin AA, et al. (2008) Weight loss treatment influences untreated spouses and the home environment: Evidence of a ripple effect. *International Journal of Obesity* 32: 1678–84.

28. Hatfield E, Cacioppo JT, Rapson RL (1994) *Emotional contagion*. (Cambridge: Cambridge University Press).

29. Collins R. (2004) *Interaction ritual chains*. (Princeton: Princeton University Press).

30. Trivers R. (1971) The evolution of reciprocal altruism. *Q. Rev. Biol.* 46: 35–57.

31. Keser C, van Winden F. (2000) Conditional cooperation and voluntary contributions to public goods. *Scand J Econ* 102: 23–39.

32. Kocher MG, et al. (2008) Conditional cooperation on three continents. *Economics Letters* 101: 175–178.

33. Fehr E, Fischbacher U (2004) Social norms and human cooperation. *Trends in Cog Sciences* 8: 185–190.

34. Weber JM, Murnighan JK (2008) Suckers or saviors? Consistent contributors in social dilemmas. *Journal of Personality and Social Psychology* 95: 1340–1353.

35. Nowak MA, Sigmund K (1998) The dynamics of indirect reciprocity. *J. Theor. Biol.* 194: 561–574.

36. Fehr E, Gachter S (2002) Altruistic punishment in humans. *Nature* 415: 137–140.

37. Alexander RD (1987) *The biology of moral systems* (New York: Aldine de Gruyter).

38. Gintis H, Smith E, Bowles S (2001) Costly signalling and cooperation. *J. Theor. Biol.* 213: 103–119.

39. Sobel ME (1982) Asymptotic confidence intervals for indirect effects in structural equation models. *Sociological Methodology* 13: 290–312.

40. Price GR (1957) Extension of covariance selection mathematics. *Annals of Human Genetics* 21: 485–490.

41. Olson M (1965) *The logic of collective action: Public goods and the theory of groups* (Cambridge: Harvard University Press).

42. Fowler JH, Dawes CT, Christakis NA (2009) Model of genetic variation in human social networks. *PNAS* 106: 1720–1724.

43. Cesarini D. et al. (2008) Heritability of cooperative behavior in the trust game. *PNAS* 105: 3721–3726.

44. Christakis NA, Fowler JH (2009) *Connected: The surprising power of our social networks and how they shape our lives* (New York: Little Brown).

45. Carman KG (2003) Social influences and the private provision of public goods: Evidence from charitable contributions in the workplace. *Stanford Institute for Economic Policy Research Discussion Paper* 02–13.

46. Rees MA, et al. (2009) A nonsimultaneous, extended, altruistic-donor chain. *New England Journal of Medicine* 360: 1096–1101.

47. Amemiya T (1973) Regression analysis when the dependent variable is truncated normal. *Econometrica* 41: 997–1016.

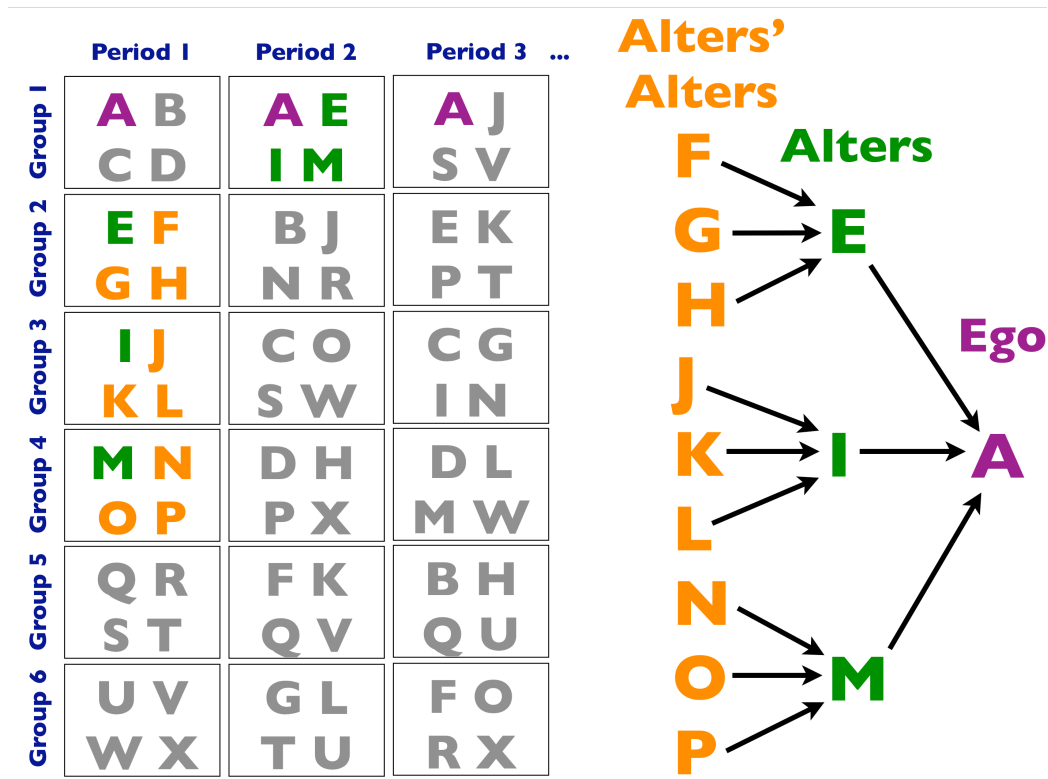

**Figure 1.** Example network drawn from the Fehr-Gaechter (*36*) public goods game experiments. Here we abstract from the numerous interactions that take place between individuals in these experiments to focus on a specific set of pathways from alters' alters to alters to egos. An "ego" is the focal subject (in this example we focus on subject A in period 3); "alters" are those subjects in the ego's group in the previous period (E, I, M in period 2). The ego has a direct network connection to alters because s/he sees each of their contributions to the public good before proceeding to the next period. "Alter's alters" are those individuals in the alters' groups in the period prior to the previous period (F, G, H, J, K, L, N, O, P in period 1). Note that the ego has *no direct network connection* to any of the alters' alters and has not seen any of their contributions. However, the ego is *indirectly connected* to the alters' alters by two degrees of separation via the alters (E, I, M in period 2). The requirement that no two subjects be placed in the same group twice guarantees that we can draw a network like this for all 24 subjects in period 3.

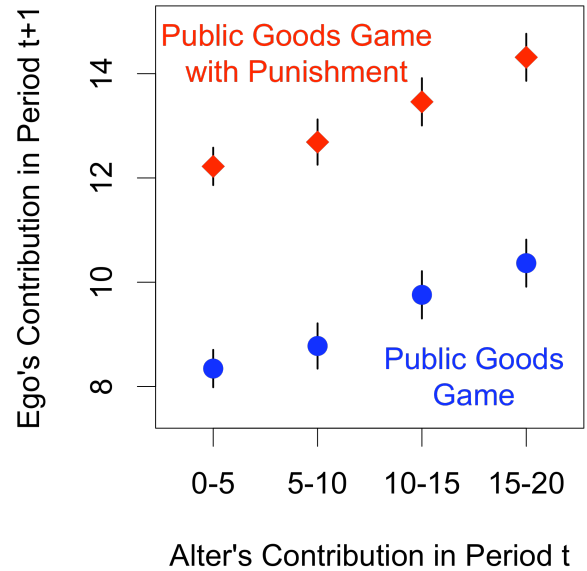

**Figure 2.** The raw data from the Fehr-Gaechter public goods game experiments (both the simple version and the version with punishment) show a relationship between alter giving in period $t$ and ego giving in period $t$+1. Individuals who gave the maximum or minimum are removed from the data to avoid floor and ceiling effects. Horizontal bars show 95% confidence intervals based on standard errors of the mean.

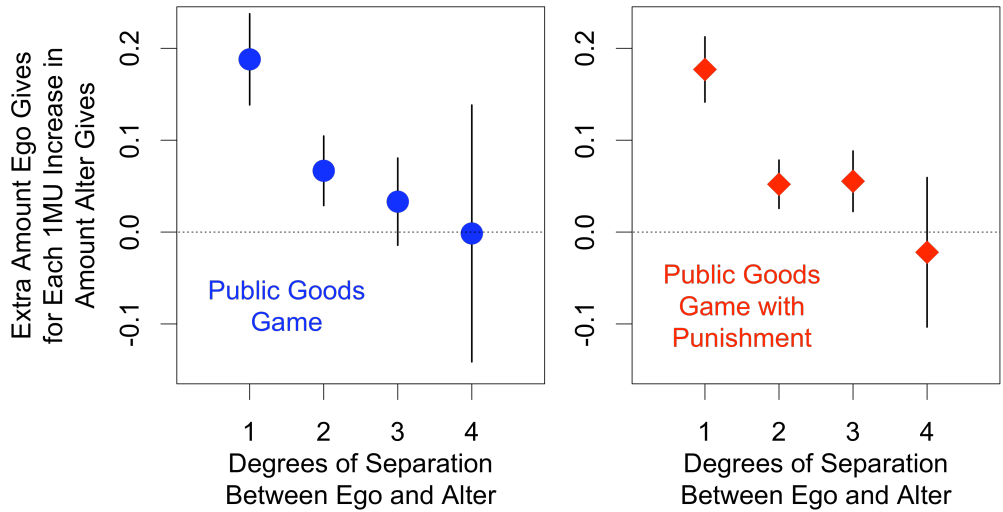

**Figure 3.** The effect of alter's contribution to the public good on ego's contribution is significant and extends up to 3 degrees of separation. For each 1MU contributed by alter, ego contributes an additional 0.19 MUs (0.18 MUs in the version with punishment) in the next period. For each 1MU contributed by alter's alter (a contribution ego did not observe), ego contributes an additional 0.07 MUs (0.05 MUs in the version with punishment) 2 periods later. And for each 1MU contributed by alter's alter's alter (3 degrees of separation), the ego contributes an additional 0.06 MUs in the public good game with punishment 3 periods later. Alters are randomly assigned to egos, and they are only assessed at the minimum degree of separation at each point in time. Estimates are from interval regressions, controlling for multiple observations of the same ego, multiple observations of the same alter, the ego's initial contribution in the period alter's contribution was observed, and period fixed effects. Horizontal bars show 95% confidence intervals.

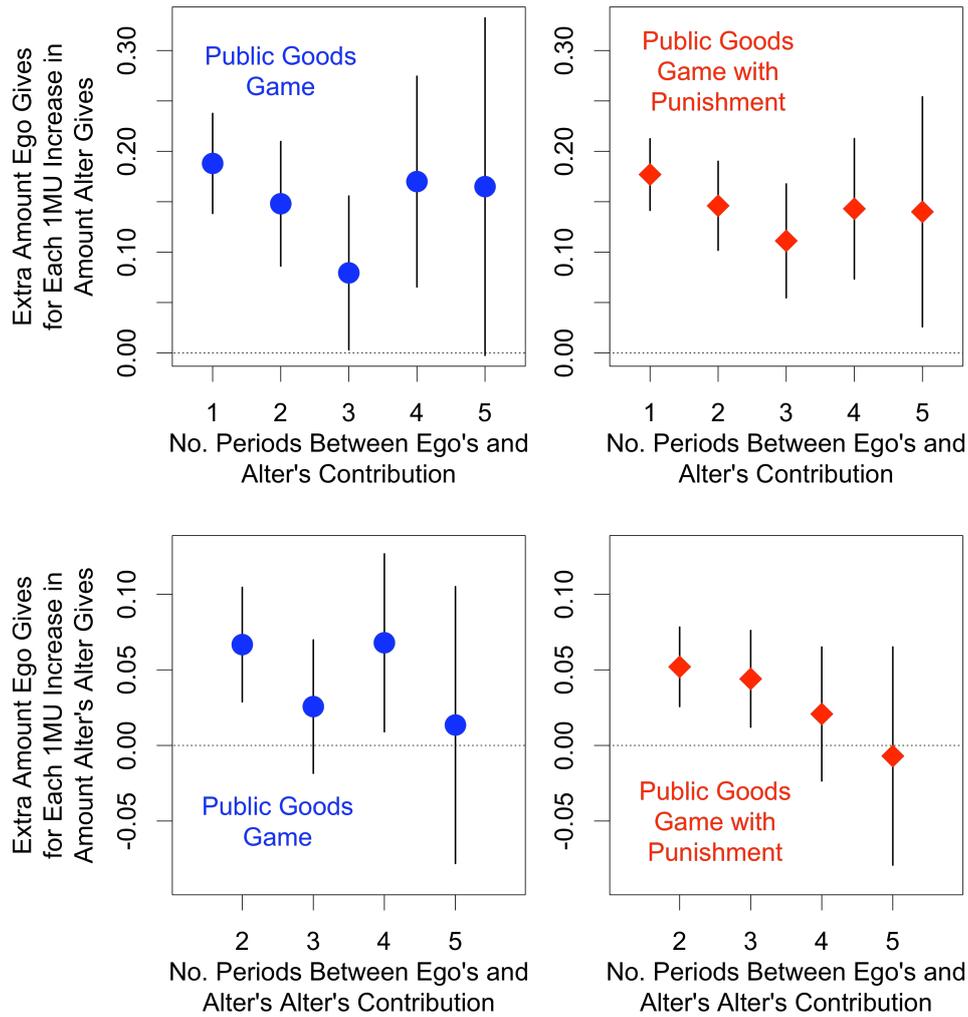

**Figure 4.** The effect of alter giving on ego giving persists beyond the initial period. Top panels show that alter significantly influences ego's behavior up to 4 periods later in the public goods game (left) and 5 periods in the public goods game with punishment (right). Bottom panels show that alter's alter (2 degrees of separation) significantly influences ego's behavior up to 4 periods later in the public goods good game and up to 3 periods later in the public goods game with punishment. Estimates are from interval regressions, controlling for multiple observations of the same ego, multiple observations of the same alter, the ego's initial contribution in the period alter's (or alter's alter's) contribution was made, and period fixed effects. Horizontal bars show 95% confidence intervals.

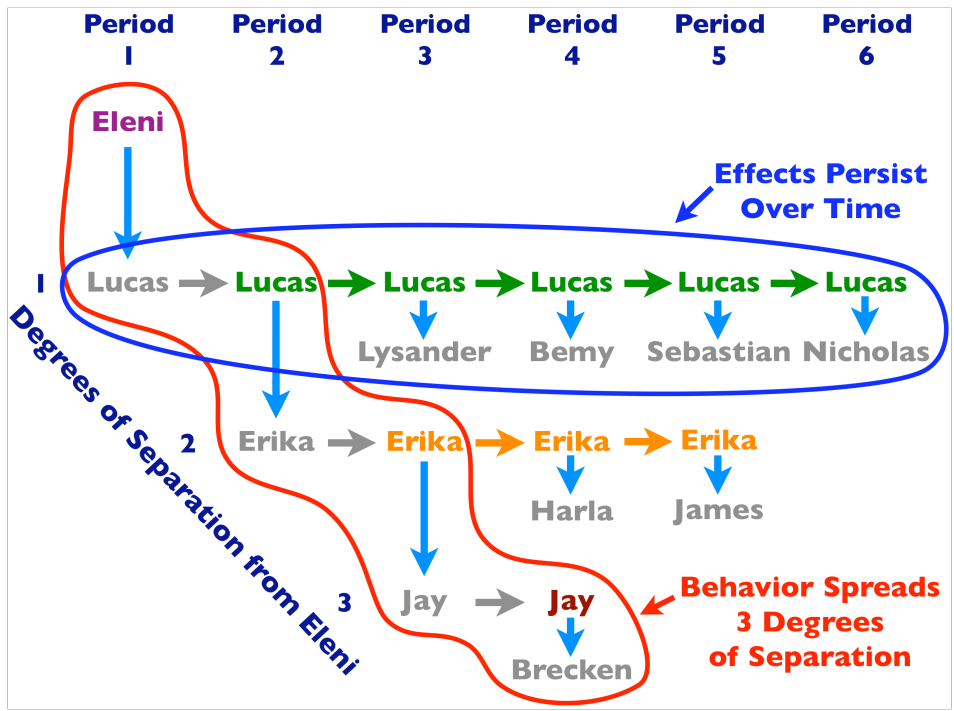

**Figure 5.** A hypothetical cascade. This diagram illustrates the difference between the spread of the inter-personal effects across individuals and the persistence of effects across time. We abstract from the numerous interactions that take place between individuals in these experiments to focus on a specific, illustrative set of pathways. Cooperative behavior spreads 3 degrees of separation: if Eleni increases her contribution to the public good, it benefits Lucas (1 degree), who gives more when paired with Erika (2 degrees) in period 2, who gives more when paired with Jay (3 degrees) in period 3, who gives more when paired with Brecken in period 4. The effects also persist over time, so that Lucas gives more when paired with Erika (period 2) and also when paired with Lysander (period 3), Bemy (period 4), Sebastian (period 5), and Nicholas (period 6). There is also persistence at 2 degrees of separation, as Erika not only gives more when paired with Jay (period 3), but also when paired with Harla (period 4) and James (period 5). All the paths in this illustrative cascade are supported by significant results in the experiments, and it is important to note that if Eleni *decreases* her initial contribution then her *uncooperative* behavior can spread and persist as well.

**Supporting Information**

# Cooperative Behavior Cascades in Human Social Networks

James H. Fowler[1*], Nicholas A. Christakis[2]

[1]Political Science Department, University of California, San Diego, La Jolla, CA 92103, USA

[2]Harvard Faculty of Arts and Sciences and Harvard Medical School, Boston, MA 02115, USA

∗ To whom correspondence should be addressed, email: jhfowler@ucsd.edu.

In Tables S1a-S1d, we show regression results used to estimate the effects shown in Fig. 3 of the main text. In Tables S1a and S2a-S2d, we show regression results used to estimate the effects shown in the top two panels of Fig. 4 of the main text. In Tables S1b and S3a-S3c, we show regression results used to estimate the effects shown in the bottom two panels of Fig. 4 of the main text. In Tables S4a-b we show regression results used to estimate the mediation effect of alter's contribution on the relationship between alter's alter's contribution and ego's contribution. In Tables S5a-b we show regression results used to estimate the effect of alter's alter's and alter's alter's alter's punishment behavior on ego's contribution.

In Tables S6a and S6b, we demonstrate that egos are not more influenced by "selfish" behavior than by "generous" behavior. In other words, alter's effect on ego does not vary for high and low contributions (increasing alter's contribution from 0MUs and 10MUs has the same effect on ego as increasing from 10MUs to 20MUs). In Table S6a, we use the *median contribution of the group* as a point of reference to divide high and low contributions, and in Table S6b we use the *ego's own contribution* as a point of reference.

In Table S7a and S7b, we explore the possibility that groups rather than specific individuals influence ego's behavior. In Table S7a, we show that alter significantly influences ego even when we include the contributions made by the other two members of the group as a control. In Table S7b, we test the influence of the other two member's contributions on alter's effect on ego by adding an interaction term to the model. The effect is significant in the public goods game but not in the public goods game with punishment, and in both cases the effect size is negligible. When other members of the group increase their contributions, it decreases the influence of alter on ego by 0.004. If we hold other group members' contributions constant at 10 each, the model suggests that an additional MU contributed by alter increases the contribution of ego in the next period by 0.160MUs, but if we increase other group members' contributions by 1MU, then an additional MU contributed by alter increases the contribution of ego in the next period by 0.156MUs. And even when others contribute maximally, alter's effect on ego remains significant ($p$=0.04).

In Table S8, we show that punishment behavior does not spread from alters to ego.

In Tables S9a-S9c, we study the effect of groups of alters rather than individual alters. These models show that the sum total of contributions by all alters in the group significantly influence ego's contribution up to two degrees of separation in the normal public goods game and up to three degrees of separation in the public goods game with punishment, mirroring the individual-level results in Tables S1a-S1c. We present these results to show that the effects are robust to specification. However, it is important to remember that the estimates in the Table S9 regressions will be downwardly biased because they do not account for censoring of individual decisions (*47*). For example, a group of alters that contributes 15+15+15=45 will have the same value as a group that contributes 20+20+5=45, but in the latter group the two individuals who gave 20 may have wanted to give more and could not because of the interval constraints of the experiment (20 was the maximum permitted contribution). In the group-based models, an observation is only counted as censored if all three alters contribute the maximum (60) so information about censoring at the individual level is lost.

In Tables S10a and S10b, we add additional lags to the model of alter's influence on ego and alter's alter's influence on ego. Note that the model results reported in Table S10 indicate that alter and alter's alter significantly influence ego in both the public goods game and the public goods game with punishment, and the effect sizes are nearly identical. However, the cost of these models is dramatically reduced sample size (an therefore the efficiency of estimation) since each additional lag eliminates one period of observations.

Finally, in Table S11, we add 235 fixed effects for each unique subject (except the baseline subject, since a constant is in the model). This method has the advantage of controlling for all fixed differences between individuals and/or sessions, but it is well-known to generate coefficients that are biased towards zero, especially when the number of fixed effects is large, as it is here. In spite of the conservative nature of this technique, we find that alter continues to have large and significant effects on ego as in the models without fixed effects.

**Table S1a: Effect of Alter's Contribution on Ego's Contribution**

| | Public Goods Game | | | Public Goods Game with Punishment | | |
|---|---|---|---|---|---|---|
| | *Coef.* | *S.E.* | *p* | *Coef.* | *S.E.* | *p* |
| *Alter's Contrib. Period t – 1* | **0.19** | **0.03** | **0.00** | **0.18** | **0.02** | **0.00** |
| *Ego's Contribution Period t – 1* | 0.98 | 0.03 | 0.00 | 0.82 | 0.02 | 0.00 |
| *Period 3* | -1.11 | 0.54 | 0.04 | -0.20 | 0.30 | 0.50 |
| *Period 4* | -1.11 | 0.52 | 0.03 | -0.62 | 0.28 | 0.03 |
| *Period 5* | -1.24 | 0.54 | 0.02 | -0.16 | 0.30 | 0.60 |
| *Period 6* | -2.64 | 0.60 | 0.00 | -0.91 | 0.37 | 0.01 |
| *Constant* | -3.89 | 0.51 | 0.00 | 2.26 | 0.38 | 0.00 |
| *Log Likelihood* | -7721 | | | -8190 | | |
| *Null Log Likelihood* | -8499 | | | -9035 | | |
| *N* | 3480 | | | 3480 | | |

*Dependent Variable:*
*Ego's Contribution in Period t*

Interval regression models of effect of alter's contribution on ego's contribution, controlling for ego's prior contribution, period fixed effects, and multiple observations of the same ego and multiple observations of the same alter using Huber-White sandwich standard errors.

**Table S1b: Effect of Alter's Alter's Contribution on Ego's Contribution**

|  | _Dependent Variable:_ | | | | | |
|  | _Ego's Contribution in Period t_ | | | | | |
|  | _Public Goods Game_ | | | _Public Goods Game with Punishment_ | | |
|  | _Coef._ | _S.E._ | _p_ | _Coef._ | _S.E._ | _p_ |
| _Alter's Alter's Contribution Period t – 2_ | **0.07** | **0.02** | **0.00** | **0.05** | **0.01** | **0.00** |
| _Ego's Contribution Period t – 2_ | 0.90 | 0.04 | 0.00 | 0.70 | 0.03 | 0.00 |
| _Period 4_ | -0.69 | 0.60 | 0.25 | -0.56 | 0.33 | 0.09 |
| _Period 5_ | -0.77 | 0.63 | 0.23 | -0.33 | 0.34 | 0.34 |
| _Period 6_ | -1.60 | 0.68 | 0.02 | -0.68 | 0.40 | 0.09 |
| _Constant_ | -4.57 | 0.60 | 0.00 | 6.49 | 0.43 | 0.00 |
| _Log Likelihood_ | -18346 | | | -19870 | | |
| _Null Log Likelihood_ | -19553 | | | -21156 | | |
| _N_ | 8316 | | | 8316 | | |

Interval regression models of effect of alter's alter's contribution on ego's contribution, controlling for ego's prior contribution, period fixed effects, and multiple observations of the same ego and multiple observations of the same alter using Huber-White sandwich standard errors.

**Table S1c: Effect of Alter's Alter's Alter's Contribution on Ego's Contribution**

| | Dependent Variable: Ego's Contribution in Period t | | | | | |
|---|---|---|---|---|---|---|
| | Public Goods Game | | | Public Goods Game with Punishment | | |
| | Coef. | S.E. | p | Coef. | S.E. | p |
| Alter's Alter's Alter's Contrib. Period t – 3 | **0.03** | **0.02** | **0.17** | **0.06** | **0.02** | **0.00** |
| Ego's Contribution Period t – 3 | 0.77 | 0.03 | 0.00 | 0.65 | 0.02 | 0.00 |
| Period 5 | -0.53 | 0.35 | 0.14 | 0.06 | 0.20 | 0.77 |
| Period 6 | -1.65 | 0.40 | 0.00 | -0.63 | 0.23 | 0.01 |
| Constant | -4.85 | 0.47 | 0.00 | 7.74 | 0.39 | 0.00 |
| Log Likelihood | | -13767 | | | -15086 | |
| Null Log Likelihood | | -14353 | | | -15796 | |
| N | | 6355 | | | 6355 | |

Interval regression models of effect of alter's alter's alter's contribution on ego's contribution, controlling for ego's prior contribution, period fixed effects, and multiple observations of the same ego and multiple observations of the same alter using Huber-White sandwich standard errors.

**Table S1d: Effect of Alter's Alter's Alter's Alter's Contribution on Ego's Contribution**

|  | Dependent Variable: Ego's Contribution in Period t | | | | | |
|  | Public Goods Game | | | Public Goods Game with Punishment | | |
|  | *Coef.* | *S.E.* | *p* | *Coef.* | *S.E.* | *p* |
|---|---|---|---|---|---|---|
| *Alter's Alter's Alter's Alter's Contrib. Period t – 4* | **0.00** | **0.07** | **0.98** | **-0.02** | **0.04** | **0.60** |
| *Ego's Contribution Period t – 4* | 0.79 | 0.08 | 0.00 | 0.51 | 0.05 | 0.00 |
| *Period 6* | 1.39 | 1.18 | 0.24 | -1.33 | 0.48 | 0.01 |
| *Constant* | -8.08 | 1.50 | 0.00 | 12.02 | 1.01 | 0.00 |
| *Log Likelihood* | | -2097 | | | -2559 | |
| *Null Log Likelihood* | | -2161 | | | -2636 | |
| *N* | | 1026 | | | 1080 | |

Interval regression models of effect of alter's alter's alter's alter's contribution on ego's contribution, controlling for ego's prior contribution, period fixed effects, and multiple observations of the same ego and multiple observations of the same alter using Huber-White sandwich standard errors.

**Table S2a: Effect of Alter's Contribution on Ego's Contribution Two Periods Later**

|  | Public Goods Game | | | Public Goods Game with Punishment | | |
| --- | --- | --- | --- | --- | --- | --- |
|  | *Coef.* | *S.E.* | *p* | *Coef.* | *S.E.* | *p* |
| *Alter's Contrib. Period t – 2* | **0.15** | **0.03** | **0.00** | **0.15** | **0.02** | **0.00** |
| *Ego's Contribution Period t – 2* | 0.90 | 0.04 | 0.00 | 0.70 | 0.03 | 0.00 |
| *Period 4* | -0.58 | 0.60 | 0.33 | -0.69 | 0.33 | 0.04 |
| *Period 5* | -0.58 | 0.64 | 0.37 | -0.55 | 0.34 | 0.11 |
| *Period 6* | -1.29 | 0.69 | 0.06 | -0.96 | 0.40 | 0.02 |
| *Constant* | -5.38 | 0.66 | 0.00 | 5.31 | 0.48 | 0.00 |
| *Log Likelihood* | | -6106 | | | -6605 | |
| *Null Log Likelihood* | | -6518 | | | -7052 | |
| *N* | | 2772 | | | 2772 | |

Dependent Variable: Ego's Contribution in Period t

Interval regression models of effect of alter's contribution on ego's contribution, controlling for ego's prior contribution, period fixed effects, and multiple observations of the same ego and multiple observations of the same alter using Huber-White sandwich standard errors.

**Table S2b: Effect of Alter's Contribution on Ego's Contribution Three Periods Later**

| | Public Goods Game | | | Public Goods Game with Punishment | | |
|---|---|---|---|---|---|---|
| | Coef. | S.E. | p | Coef. | S.E. | p |
| *Alter's Contrib. Period t – 3* | **0.08** | **0.04** | **0.04** | **0.11** | **0.03** | **0.00** |
| *Ego's Contribution Period t – 3* | 0.78 | 0.04 | 0.00 | 0.63 | 0.03 | 0.00 |
| *Period 5* | -0.64 | 0.65 | 0.33 | -0.06 | 0.35 | 0.86 |
| *Period 6* | -1.34 | 0.71 | 0.06 | -0.73 | 0.41 | 0.07 |
| *Constant* | -5.35 | 0.74 | 0.00 | 7.13 | 0.63 | 0.00 |
| *Log Likelihood* | | -4499 | | | -4926 | |
| *Null Log Likelihood* | | -4699 | | | -5169 | |
| *N* | | 2064 | | | 2064 | |

*Dependent Variable: Ego's Contribution in Period t*

Interval regression models of effect of alter's contribution on ego's contribution, controlling for ego's prior contribution, period fixed effects, and multiple observations of the same ego and multiple observations of the same alter using Huber-White sandwich standard errors.

**Table S2c: Effect of Alter's Contribution on Ego's Contribution Four Periods Later**

| | Dependent Variable: Ego's Contribution in Period t | | | | | |
|---|---|---|---|---|---|---|
| | Public Goods Game | | | Public Goods Game with Punishment | | |
| | Coef. | S.E. | p | Coef. | S.E. | p |
| Alter's Contrib. Period t – 4 | **0.17** | **0.05** | **0.00** | **0.14** | **0.04** | **0.00** |
| Ego's Contribution Period t – 4 | 0.77 | 0.06 | 0.00 | 0.63 | 0.04 | 0.00 |
| Period 6 | -1.34 | 0.78 | 0.09 | -1.03 | 0.42 | 0.02 |
| Constant | -8.09 | 1.02 | 0.00 | 7.79 | 0.73 | 0.00 |
| Log Likelihood | | -2825 | | | -3166 | |
| Null Log Likelihood | | -2928 | | | -3317 | |
| N | | 1356 | | | 1356 | |

Interval regression models of effect of alter's contribution on ego's contribution, controlling for ego's prior contribution, period fixed effects, and multiple observations of the same ego and multiple observations of the same alter using Huber-White sandwich standard errors.

**Table S2d: Effect of Alter's Contribution on Ego's Contribution Five Periods Later**

| | Public Goods Game | | | Public Goods Game with Punishment | | |
|---|---|---|---|---|---|---|
| | *Coef.* | *S.E.* | *p* | *Coef.* | *S.E.* | *p* |
| *Alter's Contrib. Period t – 5* | **0.17** | **0.09** | **0.05** | **0.14** | **0.06** | **0.02** |
| *Ego's Contribution Period t – 5* | 0.65 | 0.09 | 0.00 | 0.58 | 0.07 | 0.00 |
| *Constant* | -9.52 | 1.51 | 0.00 | 9.13 | 1.16 | 0.00 |
| *Log Likelihood* | | -1287 | | | -1500 | |
| *Null Log Likelihood* | | -1316 | | | -1544 | |
| *N* | | 648 | | | 648 | |

Dependent Variable: Ego's Contribution in Period t

Interval regression models of effect of alter's contribution on ego's contribution, controlling for ego's prior contribution, period fixed effects, and multiple observations of the same ego and multiple observations of the same alter using Huber-White sandwich standard errors.

**Table S3a: Effect of Alter's Alter's Contribution on Ego's Contribution Three Periods Later**

| | *Public Goods Game* | | | *Public Goods Game with Punishment* | | |
|---|---|---|---|---|---|---|
| | *Coef.* | *S.E.* | *p* | *Coef.* | *S.E.* | *p* |
| *Alter's Alter's Contribution Period t – 3* | **0.03** | **0.02** | **0.25** | **0.04** | **0.02** | **0.01** |
| *Ego's Contribution Period t – 3* | 0.78 | 0.04 | 0.00 | 0.64 | 0.03 | 0.00 |
| *Period 5* | -0.70 | 0.65 | 0.29 | 0.01 | 0.35 | 0.98 |
| *Period 6* | -1.46 | 0.70 | 0.04 | -0.59 | 0.41 | 0.15 |
| *Constant* | -4.81 | 0.66 | 0.00 | 7.92 | 0.54 | 0.00 |
| *Log Likelihood* | | -13501 | | | -14798 | |
| *Null Log Likelihood* | | -14098 | | | -15508 | |
| *N* | | 6192 | | | 6192 | |

*Dependent Variable:*
*Ego's Contribution in Period t*

Interval regression models of effect of alter's alter's contribution on ego's contribution, controlling for ego's prior contribution, period fixed effects, and multiple observations of the same ego and multiple observations of the same alter using Huber-White sandwich standard errors.

**Table S3b: Effect of Alter's Alter's Contribution on Ego's Contribution Four Periods Later**

| | _Dependent Variable:_ | | | | | |
| | _Ego's Contribution in Period t_ | | | | | |
| | **Public Goods Game** | | | **Public Goods Game with Punishment** | | |
| | _Coef._ | _S.E._ | _p_ | _Coef._ | _S.E._ | _p_ |
|---|---|---|---|---|---|---|
| _Alter's Alter's Contribution Period t – 4_ | **0.07** | **0.03** | **0.02** | **0.02** | **0.02** | **0.36** |
| _Ego's Contribution Period t – 4_ | 0.76 | 0.06 | 0.00 | 0.63 | 0.04 | 0.00 |
| _Period 6_ | -1.44 | 0.78 | 0.06 | -0.82 | 0.43 | 0.05 |
| _Constant_ | -6.98 | 0.90 | 0.00 | 9.32 | 0.66 | 0.00 |
| _Log Likelihood_ | | -8489 | | | -9521 | |
| _Null Log Likelihood_ | | -8785 | | | -9951 | |
| _N_ | | 4068 | | | 4068 | |

Interval regression models of effect of alter's alter's contribution on ego's contribution, controlling for ego's prior contribution, period fixed effects, and multiple observations of the same ego and multiple observations of the same alter using Huber-White sandwich standard errors.

**Table S3c: Effect of Alter's Alter's Contribution on Ego's Contribution Five Periods Later**

| | Dependent Variable: Ego's Contribution in Period t | | | | | |
| | Public Goods Game | | | Public Goods Game with Punishment | | |
| | Coef. | S.E. | p | Coef. | S.E. | p |
|---|---|---|---|---|---|---|
| *Alter's Alter's Contribution Period t – 5* | **0.01** | **0.05** | **0.77** | **-0.01** | **0.04** | **0.85** |
| *Ego's Contribution Period t – 5* | 0.65 | 0.09 | 0.00 | 0.58 | 0.07 | 0.00 |
| *Constant* | -8.02 | 1.31 | 0.00 | 10.98 | 1.05 | 0.00 |
| *Log Likelihood* | | -3867 | | | -4508 | |
| *Null Log Likelihood* | | -3948 | | | -4630 | |
| *N* | | 1944 | | | 1944 | |

Interval regression models of effect of alter's alter's contribution on ego's contribution, controlling for ego's prior contribution, period fixed effects, and multiple observations of the same ego and multiple observations of the same alter using Huber-White sandwich standard errors.

**Table S4a: Mediation Analysis, Public Goods Game**

| | Ego's Contribution in Period t | | | Alter's Contribution in Period t-1 | | | Ego's Contribution in Period t | | |
|---|---|---|---|---|---|---|---|---|---|
| | *Coef.* | *S.E.* | *p* | *Coef.* | *S.E.* | *p* | *Coef.* | *S.E.* | *p* |
| *Alter's Alter's Contrib. Period t – 2* | **0.07** | **0.02** | **0.00** | **0.28** | **0.03** | **0.00** | 0.03 | 0.02 | 0.15 |
| *Alter's Contribution Period t – 1* | --- | --- | --- | --- | --- | --- | **0.24** | **0.03** | **0.00** |
| *Ego's Contribution Period t – 2* | 0.90 | 0.04 | 0.00 | 0.03 | 0.02 | 0.11 | 0.90 | 0.04 | 0.00 |
| *Period 4* | -0.69 | 0.60 | 0.25 | -2.17 | 0.67 | 0.00 | -0.38 | 0.59 | 0.52 |
| *Period 5* | -0.77 | 0.63 | 0.23 | -3.42 | 0.67 | 0.00 | -0.29 | 0.63 | 0.64 |
| *Period 6* | -1.60 | 0.68 | 0.02 | -4.49 | 0.72 | 0.00 | -1.03 | 0.68 | 0.13 |
| *Constant* | -4.57 | 0.60 | 0.00 | 4.51 | 0.62 | 0.00 | -6.27 | 0.64 | 0.00 |
| *Log Likelihood* | -18346 | | | -20756 | | | -18257 | | |
| *Null Log Likelihood* | -19553 | | | -20970 | | | -19553 | | |
| *N* | 8316 | | | 8316 | | | 8316 | | |

Interval regression models of ego and alter contributions, controlling for multiple observations of the same ego and multiple observations of the same alter using Huber-White sandwich standard errors. The results in the first model show that the size of the direct effect of alter's alter on ego is 0.07 (95% C.I. 0.03 to 0.10). We can use the results of the second and third model to calculate the size of the indirect effect of alter's alter on ego that is mediated by alter, which is 0.07 (95% C.I. 0.05 to 0.09).

**Table S4b: Mediation Analysis, Public Goods Game with Punishment**

| | Ego's Contribution in Period t | | | Alter's Contribution in Period t – 1 | | | Ego's Contribution in Period t | | |
|---|---|---|---|---|---|---|---|---|---|
| | *Coef.* | *S.E.* | *p* | *Coef.* | *S.E.* | *p* | *Coef.* | *S.E.* | *p* |
| *Alter's Alter's Contrib. Period t – 2* | **0.05** | **0.01** | **0.00** | **0.19** | **0.03** | **0.00** | 0.03 | 0.01 | 0.04 |
| *Alter's Contribution Period t – 1* | | | | | | | **0.17** | **0.02** | **0.00** |
| *Ego's Contribution Period t – 2* | 0.70 | 0.03 | 0.00 | 0.01 | 0.01 | 0.38 | 0.70 | 0.03 | 0.00 |
| *Period 4* | -0.56 | 0.33 | 0.09 | 0.91 | 0.38 | 0.02 | -0.70 | 0.33 | 0.03 |
| *Period 5* | -0.33 | 0.34 | 0.34 | 1.20 | 0.38 | 0.00 | -0.51 | 0.34 | 0.13 |
| *Period 6* | -0.68 | 0.40 | 0.09 | 2.46 | 0.41 | 0.00 | -1.00 | 0.40 | 0.01 |
| *Constant* | 6.49 | 0.43 | 0.00 | 12.17 | 0.44 | 0.00 | 4.40 | 0.50 | 0.00 |
| *Log Likelihood* | -19870 | | | -21631 | | | -19791 | | |
| *Null Log Likelihood* | -21156 | | | -21811 | | | -21156 | | |
| *N* | 8316 | | | 8316 | | | 8316 | | |

Interval regression models of ego and alter contributions, controlling for multiple observations of the same ego and multiple observations of the same alter using Huber-White sandwich standard errors. The results in the first model show that the size of the direct effect of alter's alter on ego is 0.05 (95% C.I. 0.03 to 0.08). We can use the results of the second and third model to calculate the size of the indirect effect of alter's alter on ego that is mediated by alter, which is 0.03 (95% C.I. 0.02 to 0.05).

**Table S5a: Effect of Alter's Received Punishment on Ego's Contribution Two Rounds Later**

|  | _Dependent Variable:_ | | |
|  | _Ego's Contribution in Period t_ | | |
|  | _Coef._ | _S.E._ | _p_ |
| --- | --- | --- | --- |
| _Punishment Rec'd by Alter in Period t – 2_ | **0.13** | **0.05** | **0.02** |
| _Alter's Contribution in Period t – 2_ | 0.18 | 0.03 | 0.00 |
| _Punishment Rec'd by Ego in Period t – 2_ | 0.45 | 0.07 | 0.00 |
| _Ego's Contribution in Period t – 2_ | 0.86 | 0.03 | 0.00 |
| _Period 4_ | -1.14 | 0.33 | 0.00 |
| _Period 5_ | -0.98 | 0.34 | 0.00 |
| _Period 6_ | -1.45 | 0.40 | 0.00 |
| _Constant_ | 1.82 | 0.62 | 0.00 |
| _Log Likelihood_ | | -6565 | |
| _Null Log Likelihood_ | | -7052 | |
| _N_ | | 2772 | |

Interval regression models of effect of alter's received punishment on ego's contribution, controlling for ego's prior received punishment and ego's and alter's prior contribution, period fixed effects, and multiple observations of the same ego and multiple observations of the same alter using Huber-White sandwich standard errors.

**Table S5b: Effect of Alter's Alter's Received Punishment on Ego's Contribution Three Rounds Later**

|  | Dependent Variable: | | |
|  | Ego's Contribution in Period $t$ | | |
|  | Coef. | S.E. | p |
| --- | --- | --- | --- |
| *Punishment Rec'd by Alter's Alter in Period $t - 3$* | **0.09** | **0.07** | **0.25** |
| *Alter's Alter's Contribution in Period $t - 3$* | 0.02 | 0.04 | 0.64 |
| *Punishment Rec'd by Ego in Period $t - 3$* | 0.30 | 0.08 | 0.00 |
| *Ego's Contribution in Period $t - 3$* | 0.75 | 0.05 | 0.00 |
| *Period 5* | -0.08 | 0.41 | 0.84 |
| *Period 6* | -0.74 | 0.45 | 0.10 |
| *Constant* | 6.04 | 0.90 | 0.00 |
| *Log Likelihood* |  | -5004 |  |
| *Null Log Likelihood* |  | -5244 |  |
| *N* |  | 2118 |  |

Interval regression models of effect of alter's alter's received punishment on ego's contribution, controlling for ego's prior received punishment and ego's and alter's alter's prior contribution, period fixed effects, and multiple observations of the same ego and multiple observations of the same alter's alter using Huber-White sandwich standard errors.

**Table S6a: Effect of Alter's Contribution on Ego's Contribution is Similar Regardless of Whether Alter's Contribution is High or Low**

| | _Dependent Variable:_ | | | | | |
| --- | --- | --- | --- | --- | --- | --- |
| | _Ego's Contribution in Period t_ | | | | | |
| | _Public Goods Game_ | | | _Public Goods Game with Punishment_ | | |
| | _Coef._ | _S.E._ | _p_ | _Coef._ | _S.E._ | _p_ |
| _Alter's Contrib. Period t – 1 X Alter's Contrib. Period t – 1 > Median_ | **-0.23** | **0.14** | **0.10** | **0.02** | **0.10** | **0.87** |
| _Alter's Contrib. Period t – 1 > Median_ | 3.23 | 0.88 | 0.00 | -0.28 | 1.86 | 0.88 |
| _Alter's Contrib. Period t – 1_ | 0.23 | 0.13 | 0.07 | 0.17 | 0.03 | 0.00 |
| _Ego's Contribution Period t – 1_ | 0.98 | 0.03 | 0.00 | 0.82 | 0.02 | 0.00 |
| _Period 3_ | -1.09 | 0.54 | 0.04 | -0.20 | 0.30 | 0.51 |
| _Period 4_ | -1.06 | 0.52 | 0.04 | -0.62 | 0.28 | 0.03 |
| _Period 5_ | -1.18 | 0.54 | 0.03 | -0.15 | 0.30 | 0.62 |
| _Period 6_ | -2.53 | 0.60 | 0.00 | -0.91 | 0.37 | 0.01 |
| _Constant_ | -4.28 | 0.55 | 0.00 | 2.28 | 0.42 | 0.00 |
| _Log Likelihood_ | -7714 | | | -8190 | | |
| _Null Log Likelihood_ | -8499 | | | -9035 | | |
| _N_ | 3480 | | | 3480 | | |

Interval regression models of effect of alter's contribution on ego's contribution, controlling for ego's prior contribution, period fixed effects, and multiple observations of the same ego and multiple observations of the same alter using Huber-White sandwich standard errors. Interaction term shows the differential effect of being a higher-than-median (within the group) contributor. The interaction term is not significant, suggesting the ego does not pay more attention to higher-than-median contributors.

**Table S6b: Effect of Alter's Contribution on Ego's Contribution is Similar Regardless of Whether Alter's Contribution is Above or Below Ego's Contribution**

| | Dependent Variable: | | | | | |
| | Ego's Contribution in Period t | | | | | |
| | Public Goods Game | | | Public Goods Game with Punishment | | |
| | Coef. | S.E. | p | Coef. | S.E. | p |
|---|---|---|---|---|---|---|
| *Alter's Contrib. Period t – 1 X* | | | | | | |
| *Ego's Contrib. > Alter's Contrib. Period t – 1* | **-0.11** | **0.07** | **0.11** | **-0.02** | **0.05** | **0.70** |
| *Ego's Contrib. > Alter's Contrib. Period t – 1* | 1.83 | 0.69 | 0.00 | 0.51 | 0.77 | 0.51 |
| *Alter's Contrib. Period t – 1* | 0.26 | 0.04 | 0.00 | 0.20 | 0.04 | 0.00 |
| *Ego's Contribution Period t – 1* | 0.93 | 0.04 | 0.00 | 0.80 | 0.03 | 0.00 |
| *Period 3* | -1.10 | 0.54 | 0.04 | -0.21 | 0.30 | 0.49 |
| *Period 4* | -1.13 | 0.52 | 0.03 | -0.62 | 0.28 | 0.03 |
| *Period 5* | -1.28 | 0.54 | 0.02 | -0.16 | 0.30 | 0.60 |
| *Period 6* | -2.64 | 0.60 | 0.00 | -0.91 | 0.37 | 0.01 |
| *Constant* | -4.67 | 0.60 | 0.00 | 1.93 | 0.61 | 0.00 |
| *Log Likelihood* | | -7717 | | | -8190 | |
| *Null Log Likelihood* | | -8499 | | | -9035 | |
| *N* | | 3480 | | | 3480 | |

Interval regression models of effect of alter's contribution on ego's contribution, controlling for ego's prior contribution, period fixed effects, and multiple observations of the same ego and multiple observations of the same alter using Huber-White sandwich standard errors. Interaction term shows the differential effect of being a higher-than-median (within the group) contributor. The interaction term is not significant, suggesting the ego does not pay more attention to alters who contribute more than they did.

**Table S7a: Effect of Alter's Contribution on Ego's Contribution, Controlling for Others' Contributions**

| | *Dependent Variable:* | | | | | |
| | *Ego's Contribution in Period t* | | | | | |
| | **Public Goods Game** | | | **Public Goods Game with Punishment** | | |
| | *Coef.* | *S.E.* | *p* | *Coef.* | *S.E.* | *p* |
|---|---|---|---|---|---|---|
| *Alter's Contrib. Period t – 1* | **0.17** | **0.03** | **0.00** | **0.17** | **0.02** | **0.00** |
| *Ego's Contribution Period t – 1* | 0.95 | 0.03 | 0.00 | 0.81 | 0.02 | 0.00 |
| *Other's Contribution Period t – 1* | 0.17 | 0.03 | 0.00 | 0.17 | 0.01 | 0.00 |
| *Period 3* | -0.77 | 0.53 | 0.15 | -0.61 | 0.29 | 0.03 |
| *Period 4* | -0.34 | 0.51 | 0.51 | -1.34 | 0.28 | 0.00 |
| *Period 5* | -0.22 | 0.54 | 0.56 | -1.01 | 0.29 | 0.00 |
| *Period 6* | -1.47 | 0.61 | 0.59 | -2.06 | 0.37 | 0.00 |
| *Constant* | -6.70 | 0.59 | 0.00 | -1.90 | 0.44 | 0.00 |
| *Log Likelihood* | | -7673 | | | -8101 | |
| *Null Log Likelihood* | | -8499 | | | -9035 | |
| *N* | | 3480 | | | 3480 | |

Interval regression models of effect of alter's contribution on ego's contribution, controlling for ego's prior contribution, period fixed effects, and multiple observations of the same ego and multiple observations of the same alter using Huber-White sandwich standard errors.

## Table S7b: Effect of Interaction Between Alter's Contribution and Others' Contributions on Ego's Contribution

| | Dependent Variable: Ego's Contribution in Period t | | | | | |
|---|---|---|---|---|---|---|
| | Public Goods Game | | | Public Goods Game with Punishment | | |
| | Coef. | S.E. | p | Coef. | S.E. | p |
| Alter's Contrib. Period t – 1 X Other's Contribution Period t – 1 | **-0.004** | **0.002** | **0.05** | **-0.001** | **0.002** | **0.49** |
| Alter's Contrib. Period t – 1 | 0.24 | 0.04 | 0.00 | 0.21 | 0.06 | 0.00 |
| Other's Contribution Period t – 1 | 0.20 | 0.02 | 0.00 | 0.19 | 0.03 | 0.00 |
| Ego's Contribution Period t – 1 | 0.95 | 0.03 | 0.00 | 0.81 | 0.02 | 0.00 |
| Period 3 | -0.81 | 0.53 | 0.13 | -0.63 | 0.29 | 0.03 |
| Period 4 | -0.38 | 0.51 | 0.45 | -1.34 | 0.28 | 0.00 |
| Period 5 | -0.25 | 0.54 | 0.64 | -1.01 | 0.29 | 0.00 |
| Period 6 | -1.48 | 0.61 | 0.02 | -2.05 | 0.37 | 0.00 |
| Constant | -7.20 | 0.62 | 0.00 | -2.45 | 0.81 | 0.00 |
| Log Likelihood | -7671 | | | -8101 | | |
| Null Log Likelihood | -8499 | | | -9035 | | |
| N | 3480 | | | 3480 | | |

Interval regression models of effect of an interaction between alter's contribution and others' contributions on ego's contribution, controlling for alter's contribution, others' contributions, ego's prior contribution, period fixed effects, and multiple observations of the same ego and multiple observations of the same alter using Huber-White sandwich standard errors.

**Table S8: Effect of Alter's Punishment Behavior on Ego's Punishment Behavior**

| | *Dependent Variable: Punishment in the Public Goods Game with Punishment* | | |
|---|---|---|---|
| | *Coef.* | *S.E.* | *p* |
| *Alters' Punishments Directed at Ego in Period t – 1* | **0.01** | **0.04** | **0.83** |
| *Ego's Punishments in Period t – 1* | 0.72 | 0.06 | 0.00 |
| *Period 3* | 0.84 | 0.36 | 0.02 |
| *Period 4* | 0.50 | 0.31 | 0.10 |
| *Period 5* | 0.21 | 0.33 | 0.54 |
| *Period 6* | 0.13 | 0.30 | 0.67 |
| *Constant* | -0.92 | 0.31 | 0.00 |
| *Log Likelihood* | | -2150 | |
| *Null Log Likelihood* | | -2306 | |
| *N* | | 1160 | |

Interval regression models of effect of alters' punishments of ego on ego's punishments of others in the next round, controlling for ego's prior punishment behavior, period fixed effects, and multiple observations of the same ego using Huber-White sandwich standard errors.

**Table S9a: Effect of Alter Group's Contribution on Ego's Contribution**

|  | _Dependent Variable:_ | | | | | |
|  | _Ego's Contribution in Period t_ | | | | | |
|  | _Public Goods Game_ | | | _Public Goods Game with Punishment_ | | |
|  | _Coef._ | _S.E._ | _p_ | _Coef._ | _S.E._ | _p_ |
| _Alter Group's Contrib. Period t – 1_ | **0.09** | **0.01** | **0.00** | **0.12** | **0.01** | **0.00** |
| _Ego's Contribution Period t – 1_ | 0.56 | 0.04 | 0.00 | 0.61 | 0.04 | 0.00 |
| _Period 3_ | -0.45 | 0.57 | 0.43 | -0.31 | 0.39 | 0.42 |
| _Period 4_ | -0.40 | 0.52 | 0.36 | -0.81 | 0.30 | 0.01 |
| _Period 5_ | -0.03 | 0.49 | 0.95 | -0.64 | 0.38 | 0.10 |
| _Period 6_ | -0.69 | 0.49 | 0.16 | -1.48 | 0.42 | 0.00 |
| _Constant_ | 0.44 | 0.51 | 0.40 | 1.56 | 0.38 | 0.00 |
| _Log Likelihood_ | -3622 | | | -3249 | | |
| _Null Log Likelihood_ | -3927 | | | -3583 | | |
| _N_ | 1160 | | | 1160 | | |

Interval regression models of effect of alter group's total contributions on ego's contribution, controlling for ego's prior contribution, period fixed effects, and multiple observations of the same ego and multiple observations of the same alter using Huber-White sandwich standard errors.

**Table S9b: Effect of Alter's Alter Group's Contribution on Ego's Contribution**

| | Dependent Variable: Ego's Contribution in Period t | | | | | |
|---|---|---|---|---|---|---|
| | Public Goods Game | | | Public Goods Game with Punishment | | |
| | Coef. | S.E. | p | Coef. | S.E. | p |
| *Alter's Alter Group's Contribution Period t – 2* | **0.07** | **0.03** | **0.01** | **0.09** | **0.03** | **0.00** |
| *Ego's Contribution Period t – 2* | 0.48 | 0.04 | 0.00 | 0.49 | 0.05 | 0.00 |
| *Period 4* | -0.34 | 0.46 | 0.46 | -0.55 | 0.31 | 0.07 |
| *Period 5* | 0.05 | 0.61 | 0.93 | -0.67 | 0.50 | 0.18 |
| *Period 6* | -0.21 | 0.61 | 0.73 | -1.18 | 0.55 | 0.03 |
| *Constant* | 0.48 | 0.84 | 0.57 | 5.28 | 1.11 | 0.00 |
| *Log Likelihood* | -2959 | | | -2682 | | |
| *Null Log Likelihood* | -3109 | | | -2837 | | |
| *N* | 924 | | | 924 | | |

Interval regression models of effect of alter's alter group's total contributions on ego's contribution, controlling for ego's prior contribution, period fixed effects, and multiple observations of the same ego and multiple observations of the same alter using Huber-White sandwich standard errors.

**Table S9c: Effect of Alter's Alter's Alter Group's Contribution on Ego's Contribution**

| | Dependent Variable: | | | | | |
|---|---|---|---|---|---|---|
| | **Ego's Contribution in Period t** | | | | | |
| | **Public Goods Game** | | | **Public Goods Game with Punishment** | | |
| | Coef. | S.E. | p | Coef. | S.E. | p |
| Alter's Alter's Alter Group's Contribution | | | | | | |
| Period t – 3 | **0.02** | **0.03** | **0.47** | **0.11** | **0.03** | **0.00** |
| Ego's Contribution Period t – 3 | 0.39 | 0.05 | 0.00 | 0.43 | 0.05 | 0.00 |
| Period 5 | -0.06 | 0.45 | 0.90 | -0.39 | 0.33 | 0.24 |
| Period 6 | -0.39 | 0.57 | 0.50 | -1.28 | 0.45 | 0.00 |
| Constant | 1.61 | 0.97 | 0.10 | 5.75 | 1.26 | 0.00 |
| Log Likelihood | | -2228 | | | -2031 | |
| Null Log Likelihood | | -2299 | | | -2116 | |
| N | | 688 | | | 688 | |

Interval regression models of effect of alter's alter's alter group's total contributions on ego's contribution, controlling for ego's prior contribution, period fixed effects, and multiple observations of the same ego and multiple observations of the same alter using Huber-White sandwich standard errors.

**Table S10a: Effect of Alter's Contribution on Ego's Contribution in a Model That Includes Additional Lags of Ego's Behavior**

| | *Dependent Variable:* | | | | | |
| | *Ego's Contribution in Period t* | | | | | |
| | *Public Goods Game* | | | *Public Goods Game with Punishment* | | |
| | *Coef.* | *S.E.* | *p* | *Coef.* | *S.E.* | *p* |
|---|---|---|---|---|---|---|
| *Alter's Contrib. Period t – 1* | **0.23** | **0.05** | **0.00** | **0.16** | **0.04** | **0.00** |
| *Ego's Contrib. Period t – 1* | 0.80 | 0.06 | 0.00 | 0.55 | 0.07 | 0.00 |
| *Ego's Contrib. Period t – 2* | 0.32 | 0.07 | 0.00 | 0.21 | 0.07 | 0.00 |
| *Ego's Contrib. Period t – 3* | 0.03 | 0.07 | 0.62 | 0.03 | 0.06 | 0.64 |
| *Ego's Contrib. Period t – 4* | 0.18 | 0.06 | 0.00 | 0.27 | 0.06 | 0.00 |
| *Period 6* | -0.99 | 0.65 | 0.13 | -1.10 | 0.37 | 0.00 |
| *Constant* | -9.17 | 0.80 | 0.00 | -0.30 | 0.80 | 0.71 |
| *Log Likelihood* | | -2621 | | | -3027 | |
| *Null Log Likelihood* | | -2928 | | | -3317 | |
| *N* | | 1356 | | | 1356 | |

Interval regression models of effect of alter's contribution on ego's contribution, controlling for ego's prior contributions in periods t-1, t-2, t-3, and t-4, a fixed effect for period 6 (vs. period 5), and multiple observations of the same ego and multiple observations of the same alter using Huber-White sandwich standard errors.

**Table S10b: Effect of Alter's Alter's Contribution on Ego's Contribution in a Model That Includes Additional Lags of Ego's Behavior**

| | *Dependent Variable:* *Ego's Contribution in Period t* | | | | | |
| | *Public Goods Game* | | | *Public Goods Game with Punishment* | | |
| | *Coef.* | *S.E.* | *p* | *Coef.* | *S.E.* | *p* |
| --- | --- | --- | --- | --- | --- | --- |
| *Alter's Alter's Contrib. Period t – 1* | **0.07** | **0.03** | **0.03** | **0.06** | **0.02** | **0.00** |
| *Ego's Contrib. Period t – 1* | 0.70 | 0.07 | 0.00 | 0.50 | 0.06 | 0.00 |
| *Ego's Contrib. Period t – 2* | 0.22 | 0.07 | 0.00 | 0.15 | 0.06 | 0.02 |
| *Ego's Contrib. Period t – 3* | 0.25 | 0.06 | 0.00 | 0.30 | 0.06 | 0.00 |
| *Period 6* | -0.65 | 0.72 | 0.36 | -0.85 | 0.40 | 0.03 |
| *Constant* | -8.66 | 0.85 | 0.00 | 3.41 | 0.69 | 0.00 |
| *Log Likelihood* | -8187 | | | -9280 | | |
| *Null Log Likelihood* | -8785 | | | -9951 | | |
| *N* | 4068 | | | 4068 | | |

Interval regression models of effect of alter's alter's contribution on ego's contribution, controlling for ego's prior contributions in periods t-1, t-2, and t-3, a fixed effect for period 6 (vs. period 5), and multiple observations of the same ego and multiple observations of the same alter using Huber-White sandwich standard errors.

**Table S11: Effect of Alter's Contribution on Ego's Contribution in a Model That Includes Ego Fixed Effects**

| | **Dependent Variable:** | | | | | |
| | **Ego's Contribution in Period t** | | | | | |
| | **Public Goods Game** | | | **Public Goods Game with Punishment** | | |
| | Coef. | S.E. | p | Coef. | S.E. | p |
|---|---|---|---|---|---|---|
| *Alter's Contrib. Period t − 1* | **0.17** | **0.02** | **0.00** | **0.14** | **0.02** | **0.00** |
| *Period 3* | -2.41 | 0.42 | 0.00 | 1.16 | 0.24 | 0.00 |
| *Period 4* | -4.02 | 0.40 | 0.00 | 1.52 | 0.23 | 0.00 |
| *Period 5* | -5.41 | 0.43 | 0.00 | 2.40 | 0.25 | 0.00 |
| *Period 6* | -7.36 | 0.50 | 0.00 | 2.06 | 0.30 | 0.00 |
| *Constant* | 0.05 | 1.58 | 0.00 | 13.21 | 0.38 | 0.00 |
| *Log Likelihood* | -6699 | | | -7292 | | |
| *Null Log Likelihood* | -8499 | | | -9035 | | |
| *N* | 3480 | | | 3480 | | |

Interval regression models of effect of alter's contribution on ego's contribution, controlling for ego's prior contribution, period fixed effects, ego fixed effects for 235 subjects, and multiple observations of the same ego and multiple observations of the same alter using Huber-White sandwich standard errors. Coefficients on fixed effects not shown.